# Unsupervised Method for Correlated Noise Removal for Multi-wavelength Exoplanet Transit Observations

Ali Dehghan Firoozabadi[1], Alejandro Diaz[1], Patricio Rojo[2], Ismael Soto[1], Rodrigo Mahu[3], Nestor Becerra Yoma[3], and Elyar Sedaghati[4]

[1] Department of Electrical Engineering, University of Santiago of Chile, Santiago, Chile; ali.dehghanfirouzabadi@usach.cl, alejandro.diaza@usach.cl, ismael.soto@usach.cl

[2] Department of Astronomy, University of Chile, Santiago, Chile; pato@das.uchile.cl

[3] Department of Electrical Engineering, University of Chile, Santiago, Chile; rmahu@ing.uchile.cl, nbecerra@ing.uchile.cl

[4] European Southern Observatory, Garching, Germany; elyar.sedaghati@dlr.de

**Abstract**

Exoplanetary atmospheric observations require an exquisite precision in the measurement of the relative flux among wavelengths. In this paper, we aim to provide a new adaptive method to treat light curves before fitting transit parameters in order to minimize systematic effects that affect, for instance, ground-based observations of exo-atmospheres. We propose a neural-network-based method that uses a reference built from the data itself with parameters that are chosen in an unsupervised fashion. To improve the performance of proposed method, K-means clustering and Silhouette criteria are used for identifying similar wavelengths in each cluster. We also constrain under which circumstances our method improves the measurement of planetary-to-stellar radius ratio without producing significant systematic offset. We tested our method in high quality data from WASP-19b and low-quality data from GJ-1214. We succeed in providing smaller error bars for the former when using JKTEBOP, but GJ-1214 light curve was beyond the capabilities of this method to improve as it was expected from our validation tests.

*Key words:* planet–star interactions – stars: atmospheres – (stars:) planetary systems

*Online material:* color figures

## 1. Introduction

The characterization of exoplanetary atmospheres is a rapidly growing subfield of extrasolar planet studies. This characterization is usually founfd among the key scientific objectives in ambitious projects like the *James Webb Space Telescope* and new generation, extremely large telescopes. Space-based observations produced the first and many of the strongest results regarding the properties of exoatmospheres (Charbonneau et al. 2002; Iyer et al. 2016, and references therein). However, since the first ground-based detection of these atmospheres (Redfield et al. 2008), an important effort has been devoted to understanding and reducing the telluric effects on the observations from ground-based facilities. Spectrographs and images mounted on 4 m class telescopes and larger are now used to constrain exoatmospheric properties of transiting planets (e.g., Astudillo-Defru & Rojo 2013; Wyttenbach et al. 2015; Brogi et al. 2016), but, the statistical significance of these detections are borderline for most cases.

There are some known methods to attenuate atmospheric perturbations in astronomic observations. In (Lazorenko & Lazorenko 2004), the authors present a novel method for atmospheric noise filtration based on apodization of the entrance pupil and the enhanced virtual symmetry of reference stars. Furthermore, there are some tools that helps comparing (within many other uses) these methods like JKTEBOP which uses Levenberg–Marquardt optimization algorithm to find the best-fitting model for a transit star or a binary eclipse, then analyzes the error through Bootstrapping To determine robust uncertainties in the parameters.

In this paper, we propose a new method inspired on a nonlinear autoregressive exogenous model (NARX), typically used to model past noise on different engineering processes. In (Mat et al. 2010), Particle Swarm Optimization (BPSO) algorithm was used to perform model structure selection (selection of input and output delays that best explains the future values of the data). The parameter estimation of the NARX model was done using Householder Transform-based QR factorization. Results show that the NARX model was successful in estimating the model, and filtering out noise optimally. Additionally, polynomial NARX models were implemented in an adaptive controller for nonlinear active noise in (Napoli & Piroddi 2009) for the control of vibrations in a duct using linear and nonlinear actuators. The controller parameters successively got updated based on the error gradient and the residual noise. In (Asgari et al. 2015), NARX models of the start-up procedure of a heavy-duty gas turbine are constructed. The results of this modeling approach allow the set up of a simulation tool, which can be used for real-time control and sensor diagnostics of gas turbines. In another study, a multilayer perception neural-network-based control structure was introduced in (Bouchard et al. 1999) as a nonlinear active controller, with a training algorithm based on an extended back-propagation scheme.

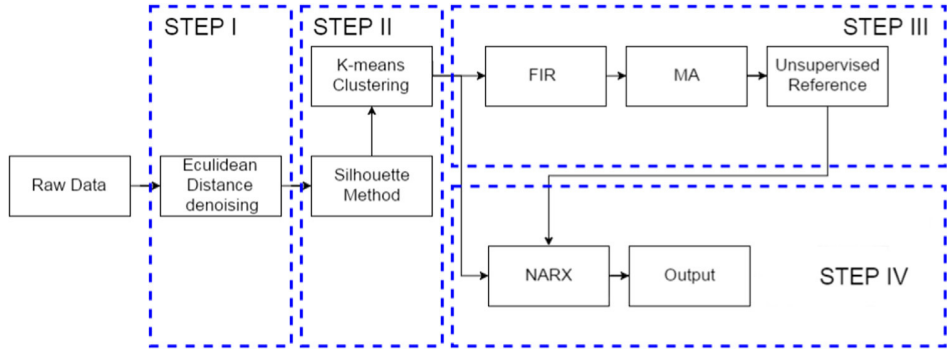

**Figure 1.** Block diagram of the proposed method.
(A color version of this figure is available in the online journal.)

NARX algorithms require a reference to be effective. We present a method to obtain such references that is based on a filtered version of the data itself and moving average. The Bandwidth to filter this data is adjusted to work with different radii ratio. A pre-processing part based on Euclidean distance is implemented to eliminate the noisy wavelengths. Also, we propose to use the K-means clustering and Silhouette method to find the similarity for wavelengths to improve the performance of the proposed method that is not sensitive to the filter parameters. We study the effect of the filter on the measurement of planet-to-star radius ratio from observations of the planets GJ-1214 and WASP-19b.

Section 2 describes the methodology and details of the proposed method. Section 3 validates the technique with data and shows the results of the proposed method on two different data sets. And finally, conclusions are presented in Section 4.

## 2. Methodology

Incoming radiation is affected by systematic effects (telluric, instrumental) that correlates at some wavelengths. The aim of the method is to eliminate this extra noise and obtain the transit light curve.

The block diagram for the proposed method is shown in Figure 1, divided into four steps: Step 1: removal of spectral regions at wavelengths that are too noisy. Step 2: cluster together spectral channels that respond similarly to the atmosphere and treat them as independent groups from that point forward. Step 3: using unsupervised method, extract a reference signal for the NARX modeling. Step 4: we use the NARX model and the reference from the previous step to eliminate the noise from the signal within the clusters.

### 2.1. STEP 1: Euclidean Distance

The Euclidean distance is the ordinary (i.e., straight-line) distance between two points in Euclidean space. The Euclidean distance between two wavelength channels can be demonstrated as

$$d(f(\lambda_n, t), f(\lambda_{n+1}, t)) = \sqrt{\sum_t (f(\lambda_n, t) - f(\lambda_{n+1}, t))^2}, \quad (1)$$

where $d(f(\lambda_n, t), f(\lambda_{n+1}, t))$ is the Euclidean distance between vectors at channels $\lambda_n$ and $\lambda_{n+1}$ and the sum is over all the epochs.

Comparison of Euclidean distances for different consecutive channels can be done visually to identify those with large value due to extreme dissimilarity. A cutoff is then chosen and channels above this value are considered too noisy and are removed form in the analysis.

### 2.2. STEP 2: Classification Process

The K-means clustering algorithm bundles similar vectors together (MacQueen 1967; Dehghan Firoozabadi & Abutalebi 2015, with an interesting application). In this case, we want to see which wavelength channels respond similarly with time. Each vector is thus a wavelength channel with its element in a different frame. The algorithm first requires the *K* centroids positions as far as possible. Afterwards, each data point is associated to the closest centroid. When all the points are associated with one of the centroids, the primary grouping is achieved. In the next step, new positions for centroids are calculated according to vectors of each cluster. Formally, the

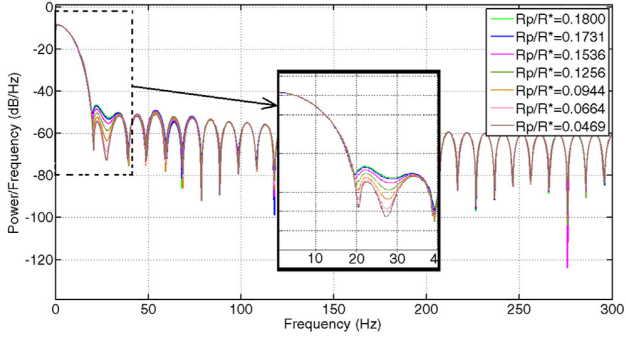

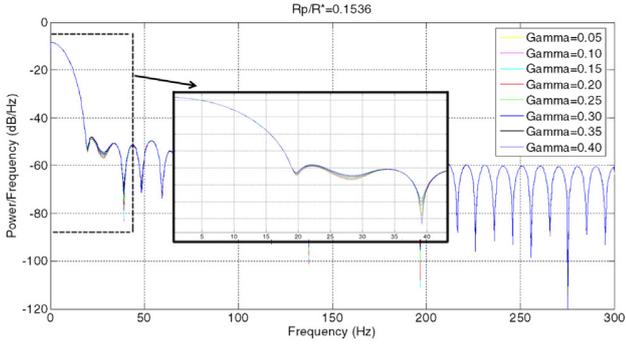

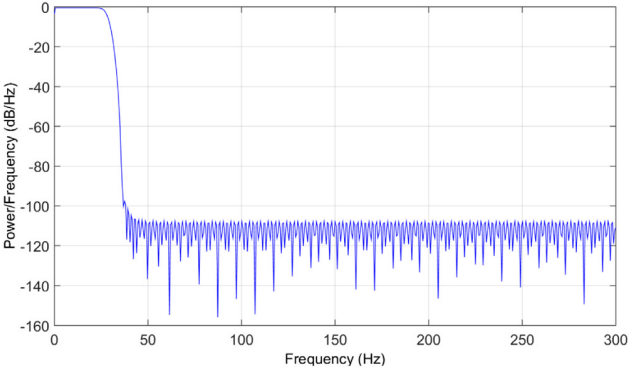

**Figure 2.** (a) Reference's periodograms at many values of radii ratio, (b) at many values of limb darkening for fixed radii ratio, and (c) calculated filter response for FIR filter with bandwidth 40 Hz.
(A color version of this figure is available in the online journal.)

**Table 1**
K-means Clusters

| Data Set | Cluster # | Wavelength Range |
|---|---|---|
| WASP-19b | 1 | 550–640 (nm) |
|  | 2 | 640–810 (nm) |
| GJ-1214 | 1 | 2.28–2.51 ($\mu$m) |
|  | 21 | 1.58–1.71 ($\mu$m) |
|  | 22 | 2.1–2.59 ($\mu$m) |
|  | 31 | 1.4–1.58 ($\mu$m) |
|  | 32 | 1.95–2.1 ($\mu$m) |

**Table 2**
Parameters for Generation Idealized Planetary Transit Light Curve for GJ-1214 ans WASP-19b

| Parameter | Description | WASP-19b | GJ-1214b |
|---|---|---|---|
| Rp/Rs | Planet-to-star radius ratio | 0.1416 | 0.1351 |
| $a$/Rs | Semimajor axis to stellar radius ratio | 3.656 | 15.31 |
| P | planet's period in days | 0.326216852 | 0.653560609 |
| i | inclination of the orbit in degrees | 80.36 | 88.17 |
| Gamma1 | Quadratic limb-darkening coefficients | 0.391 | 0.3391 |
| Gamma2 |  | 0.225 | 0.3196 |

objective is to minimize the objective function $S$:

$$S = \sum_{i=1}^{k} \sum_{\lambda_i \in \Lambda} \|\lambda_i - \mu_i\|^2, \quad (2)$$

where $\Lambda$ are all possible wavelength channels, $\mu_i$ is the mean for all wavelengths in cluster $i$, and $k$ is the numbers of wavelength. These steps are iterated until the achieved centroids stays within an error threshold. This iterative partitioning minimizes the sum, over all clusters, of the within-cluster sums of point-to-cluster-centroid distances. We used squared Euclidean distances as the scaler this minimization. In this paper we considered $10^{-3}$ as an error threshold for the centroid of clusters. Since a smaller value increases the complexity of method without increasing the accuracy.

The Silhouette method is used to find the optimal value for $K$ (Rousseeuw 1987). It provides a simple graphical alternative to find the optimal value $K$ that optimizes the clustering of data. Although different criteria can be used for measuring the dissimilarity, the Euclidean measure is widely used in this regard. We can define $a_i$ as the level of coincidence within its cluster (lower values show higher adjustment). Average dissimilarity of $\lambda_i$ to cluster $i$ is calculated by average distance of $\lambda_i$ from the centroid of cluster $i$. We also define the lowest average dissimilarity of $b_i$ for other clusters in which $\lambda_i$ is not a member of them. The cluster which has the lowest average dissimilarity is known as "neighboring cluster" for $\lambda_i$, because it is the nest best alternative cluster to fit $\lambda_i$. The Silhouette value, is defined as

$$Z(i) = \frac{b_i - a_i}{\max\{b_i, a_i\}}, \quad (3)$$

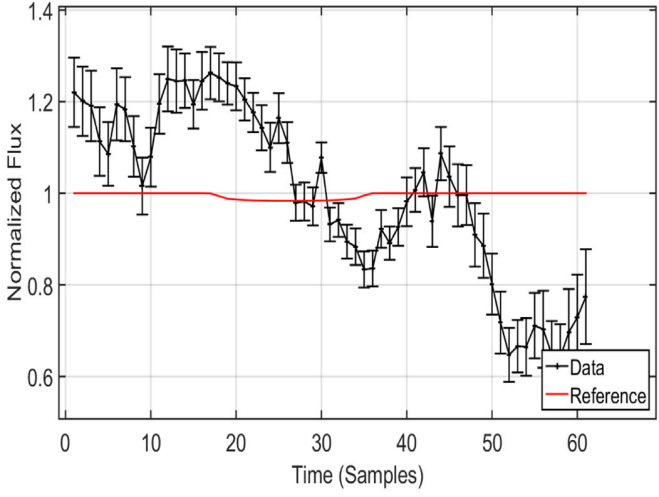
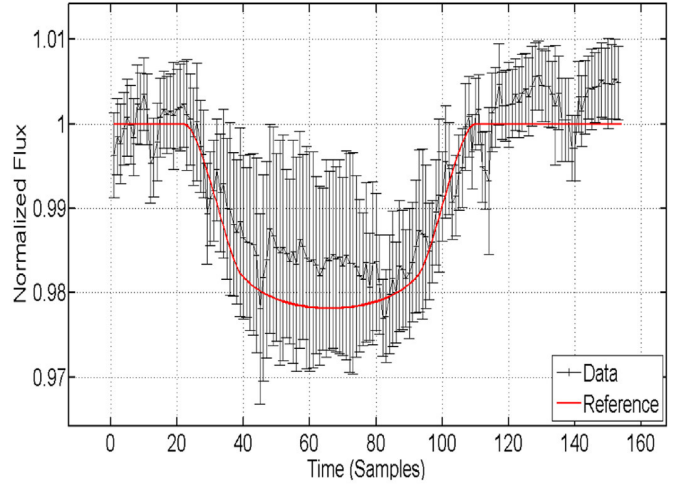
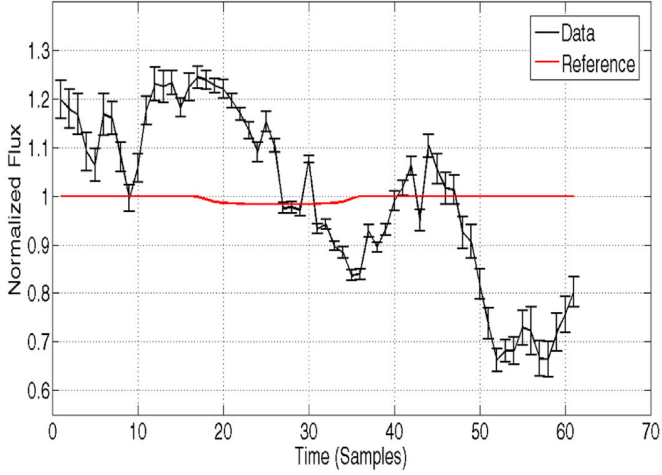
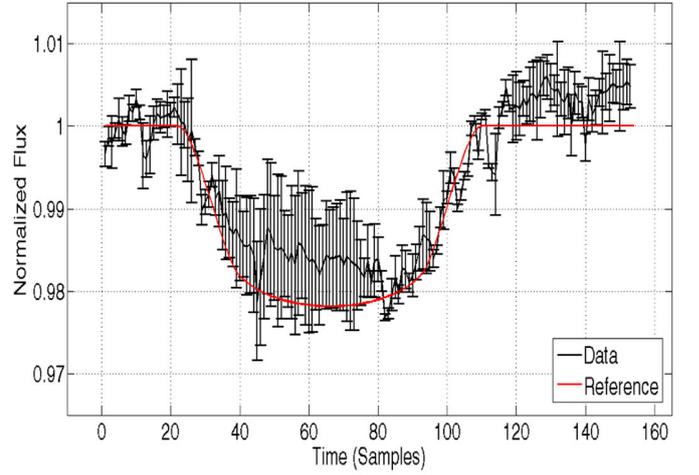

**Figure 3.** Black line represents white light from (a) GJ-1214 and (b) WASP-19b. The red line shows a reference MA02 light curve with parameters from Table 1. (c) GJ-1214 and (d) WASP-19b after removing the noisier channels with euclidean distance method.
(A color version of this figure is available in the online journal.)

where $Z(i)$ is the Silhouette value that can also be written as

$$Z(i) = \begin{cases} 1 - \dfrac{a_i}{b_i} & \text{if } a_i < b_i \\ 0 & \text{if } a_i = b_i \\ \dfrac{b_i}{a_i} - 1 & \text{if } a_i > b_i \end{cases} \quad (4)$$

If $a_i \ll b_i$, then $Z(i)$ approaches 1. Since $a_i$ shows dissimilarity of $\lambda_i$ to its cluster, low $a_i$ values show that $i$th data point has good coincidence with its cluster. In addition, a large value of $b_i$ shows that $i$th data point is not coincident with its neighbor cluster. So if $Z(i) \approx 1$, the data point has been clustered well. If $Z(i) \approx -1$, it would be better if $i$th data is transferred to the neighboring cluster.

### 2.3. STEP 3: Unsupervised Reference

We first noticed that in Fourier space, the shape of a FIR filter of order 23 (Figure 2(c)) is quite similar to the Fourier transform of an idealized planetary transit light curve (Figure 2(a)) according to (Mandel & Agol 2002, hereafter MA02). A Finite Impulse Response (FIR) filter is a filter whose impulse response (or response to any finite length input) is of finite duration, because it settles to zero in finite time. This is in contrast to infinite impulse response (IIR) filters, which may have internal feedback and may continue to respond indefinitely (usually decaying).

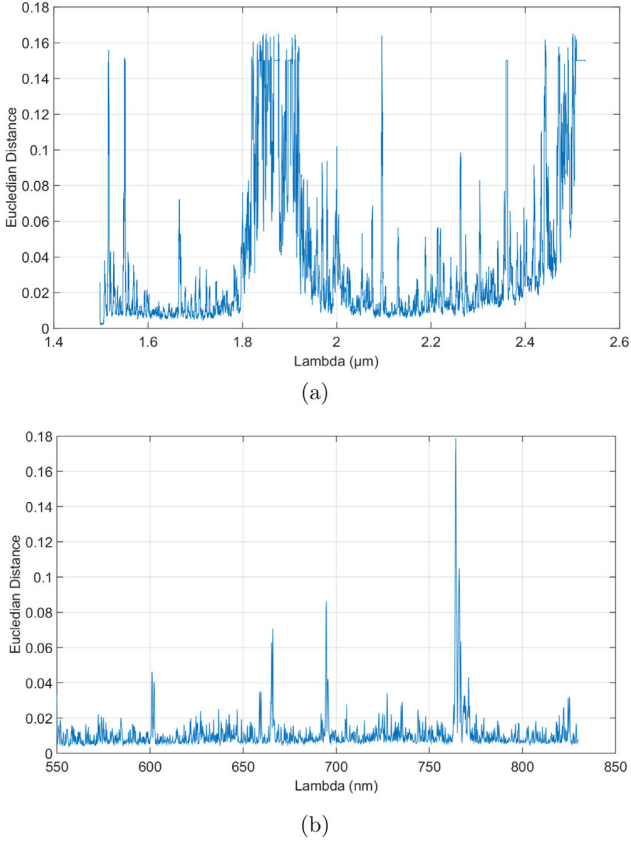

**Figure 4.** Euclidean Distance for (a) GJ-1214 and (b) WASP-19b data sets. A smaller value means similar behavior in time between continuous wavelength channels.
(A color version of this figure is available in the online journal.)

There are two main parameters for the FIR filter: Bandwidth (BW) and Rejection Band (RB). We argue that these parameters do not affect the planetary-to-stellar radius ratio measurement, which is the most relevant parameter for exoatmospheric characterization. Rejection Band indicates just how strong the removal of the high frequency components would be, while the cutoff frequency (bandwidth) preserves the information inside at lower frequency. Figure 2(a) shows the Fourier transform (FFT) for different values of radius-ratio when modeling the transit with the MA02 model and also, Figure 2(b) shows the Fourier transform (FFT) for different values of limb darkening for fixed radius-ratio. It shows that for both conditions (changes in radius-ratio and limb darkening) the difference appears mostly within the range 10–40 Hz and only changes the amplitude of FFT, not its bandwidth. Therefore, by choosing a bandwidth to the filter beyond 40 Hz (Figure 2(c)), we are not influencing the measurement of the radius ratio. In contrast, other parameters of the transit's fit might be affected.

We use Matlab's FDATool to design the filter with the following parameters: *Fpass* set to BW, *Fstop* set to 5 Hz beyond BW to avoid distortion, *Apass* set to a minimum 1dB, *Astop* set to RB, Density Factor set to a nominal 100, and *Fs* set to twice the value of *Fstop* to eliminate spatial aliasing.

*2.3.1. Moving Average*

A moving average is a sequence of arithmetic means taken over a fixed interval moved along consecutive data points from an infinite (or sufficiently large) set of data points to smooth the signal. The interval $n$ and the number of iterations are set manually, and as Section 3.2 will show, the choice for those parameter does not strongly affect the result.

In summery, in order to build the NARX reference we need to specify the following four parameters: bandwidth, rejection band (RB) for FIR filter and length of window (LW), and number of iterations (NI) for MA algorithm. There are various methods for optimizing these parameters. Simply, a Mean Square Error minimization (MMSE) of the filtered data $x_{est}$ with a fiducial MA02 curve $x_{ref}$ (with parameters taken from broadband photometry in Table 2) should suffice:

$$[BW_{opt}, RB_{opt}, LW_{opt}, NI_{opt}] = \arg\min_V \sqrt{(x_{est} - x_{ref})^2} \quad (5)$$

Note that the final result of the method is only weekly dependent on the choice of the fiducial model when trying to measure radius ratio as long as BW stays beyond 40 Hz.

*2.4. STEP 4: Denoising System*

*2.4.1. NARX System*

The neural network computes a dynamical function of its input and the reference given to achieve the output we wish to have. Details can be found an Appendix A. First, it needs to be "trained" with a set of data different from the operating set of data. Normally training uses more data to iteratively estimate the transfer function that leads from the input data to the reference given. By doing this, the neural network sets the weights of its "synapses," therefore the neural network "learns" how to obtain the desired output.

We used Bayesian regularization back-propagation network training function, that updates the weight and bias values according to Levenberg–Marquardt algorithm. It minimizes a combination of squared errors and weights, and then determines the correct combination so as to produce a network that generalizes optimally.

There are different structures of neural networks based on the number of neurons, number of layers, the existing feedback or feed forward, and the presence of hidden layers. NARX is a structure of NN with feedback. It means that there is an effect of output to use as input, but with delay. For training and validating the network we use a random sample of 10% of all wavelength channels. Of them, 70% of the epochs were used for training, 20% for validation and 10% for testing. the remaining wavelength channels (90%) are used as input to the algorithm. Number of neurons and tap delays depend on data as exemplified in nest section.

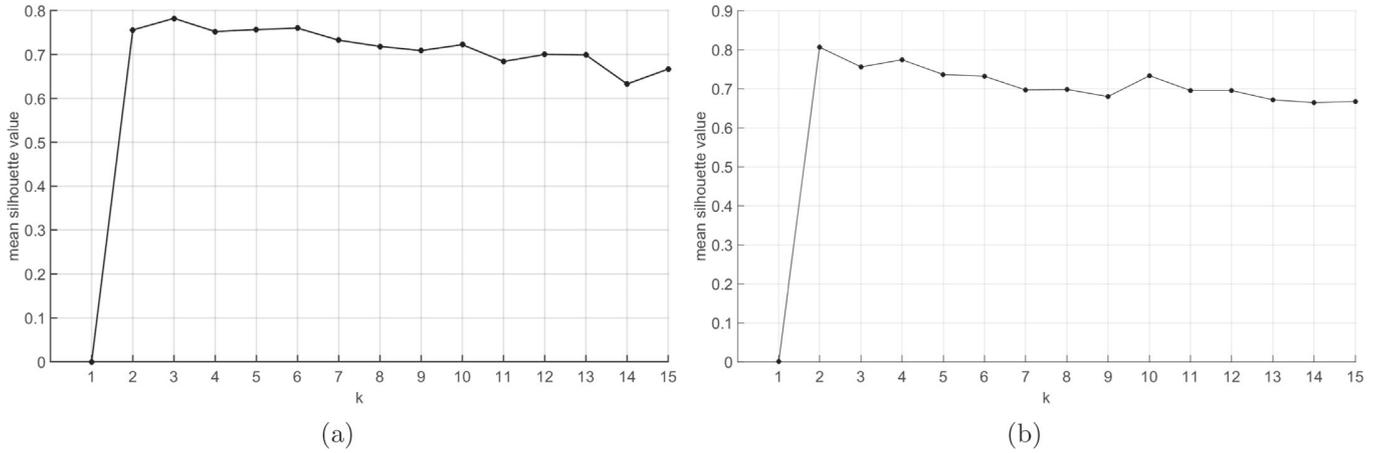

**Figure 5.** Mean Silhouette value for (a) GJ-1214 and (b)WASP-19b.

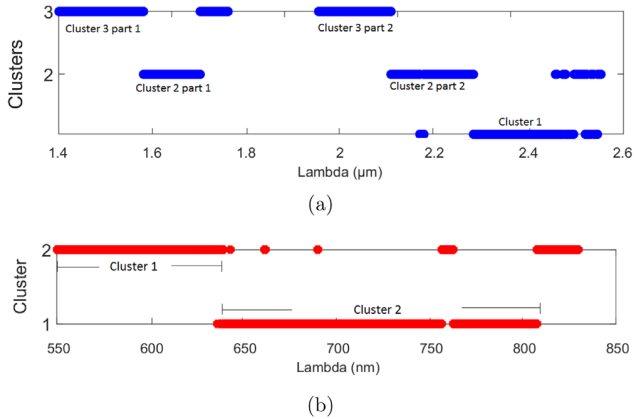

**Figure 6.** Clustering favored by K-means algorithm for (a) GJ-1214 Data set and (b) WASP-19b Data set. The vertical axis represents the class or the cluster "name" for the different wavelength channels.
(A color version of this figure is available in the online journal.)

## 3. Simulations and Results

### 3.1. Software Selection

We tested our method using two real data sets. We used MATLAB® software in every simulation step: $K$-means algorithm and silhouette method for choosing the best $K$, the Unsupervised Reference method and NARX training and operation process. In the last part we used JKTEBOP code, which uses the Levenberg–Marquardt optimization algorithm to find the best- model to the transiting planet light curve, to fit a model to the light curves (Southworth et al. 2004) and (Southworth 2014). We used task 9 in JKTEBOP software for producing these error bars using Monte Carlo simulations with the priar's bead method.

#### 3.1.1. The Data Set

The first data set is a spectral time series for the transit of the star GJ-1214, taken by SOFI at the NTT in long SCLT Red mode. The wavelength coverage was from 1.49 to 2.54 $\mu n$ in 5068 channels. These samples were taken every 185 s approximately during 3.33 hr for a total of 61 frames on 2011-08-10T00:56:22.

The second data set is a spectral transit time-series presented by (Sedaghati et al. 2015). For the planet WASP-19b on the night of 2014 November 16, it uses FORS2 at the VLT with the Grism 600RI between 0.55–0.83 $\mu$m. In addition, it uses 30″ × 10″ slits on 6 reference stars in addition to WASP-19b. The exposure time was 30s with a typical signal-to-noise ratio (S/N) of 300 at the central wavelength. Figure 3 shows the averaged white light curve for these data sets.

The original data for GJ-1214 and WASP-19b are shown in Figures 3(a) and (b) respectively. Many observed transit light curves of WASP-19b were actually affected by starspots (Huitson et al. 2013; Mancini Lombardi et al. 2013; Tregloan-Reed et al. 2013), i.e., the planet occulted starspots during its transit events. This phenomenon is color dependent and can severely contaminate the results, i.e., the estimation of the transit depth. Moreover, unoccultted starspots can also affect transit events and, again, their effects depend on wavelength, particularly stronger at blue, and on stellar activity, which can also vary according to stellar cycles; indeed, stellar activity can vary in a wide range of spatial and temporal scale.

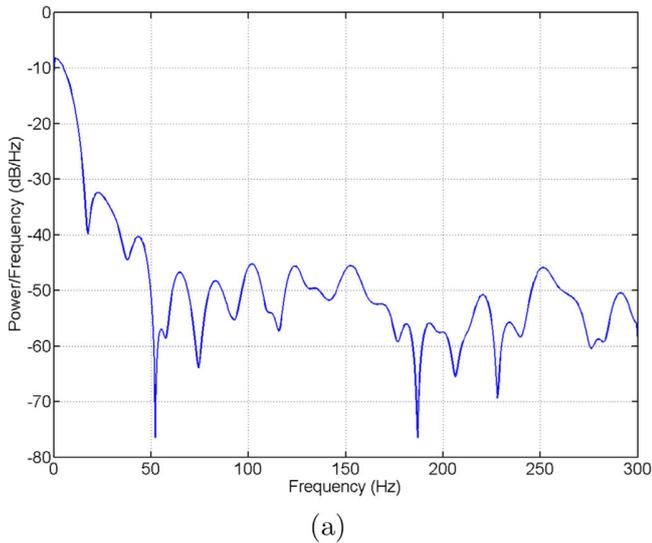
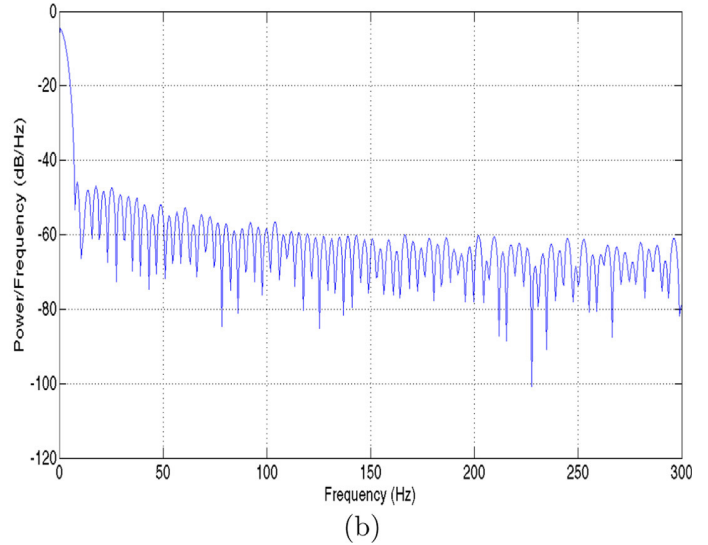

**Figure 7.** Periodogram for (a) GJ-1214 and (b) WASP-19b data set.
(A color version of this figure is available in the online journal.)

**Table 3**
Results for Optimization of FIR and MA Parameters

|  | FIR | | MA | | |
| --- | --- | --- | --- | --- | --- |
|  | BW (Hz) | RB | LW | N | MMSE |
| WASP-19b | 38 | −109 | 3 | 3 | 0.0085 |
| GJ-1214 | 43 | −117 | 3 | 4 | 0.0123 |

### 3.2. Method Testing

*Step 1.* Figure 4 shows the result of euclidean distance for two consecutive wavelength channels. We can see in Figure 4(a) (for GJ-1214) a very noisy region from $\lambda_{1.78\mu m}$ to $\lambda_{1.95\mu m}$ and $\lambda_{2.4\mu m}$ to $\lambda_{2.54\mu m}$ that correspond to the telluric region between the H and K bands. Those regions were removed. We did not remove channels for WASP-19b (Figure 4(b)) because this data is from a single spectral band. The details for all clusters are shown in Table 1.

*Step 2: Clustering.* The Silhouette method found an optimal number of clusters (*K*-value) of 3 and 2 for GJ-1214 and WASP-19b data sets, respectively (Figure 5). In each of the clusters, wavelengths will be analyzed independently (Figure 6). The optimal value of *K* can then be found by determining the *K* that maximizes the average of $\langle Z(i) \rangle_i$.

*Step 3: Unsupervised reference.* The periodogram for the white light curve of GJ-1214 and WASP-19b are shown in Figure 7. The optimization of Equation (5) yields values in Table 3 and Figure 8.

Note, the similarity of our data to a light curve without manipulation, using the look-alike FIR.

As seen in Figures 8 and 9, we plotted another figure as "Reference." This reference is an fiducial planetary transit light curve that is generated based on the parameters for WASP-19b and GJ-1214 in Table 2. These fiducial models are only shown for comparison and are never used in the algorithm.

*Step 4: NARX modeling.* For the NARX we used three neurons and five tap delays for both input, and output in WASP-19b data set and six neurons and five tap delays for both input, and output in GJ-1214 data set. We run the method a hundred times for both data sets with the purpose to cover the most initial states of NARX.

Figure 9 presents the results of the method. This figure shows the white-light planetary transit light curve in the time domain for GJ-1214 and WASP-19b respectively. To fit the resulting light curve, we use JKTEBOP's task 9, with priors taken from the literature (Table 2) and fitting only for radius ratio.

Figures 10–12 show the resulting radii ratio and wavelengths for our data set, including comparison with literature for WASP-19b that shows our proposed method has shorter error bars in comparison with (Sedaghati et al. 2015). The result for GJ-1214 is clearly non-consistent with previous results (e.g., Berta et al. 2012, which uses *Hubble Space Telescope* observations) which are over 6 sigmas at most wavelengths. In the next section, we argue that the noise of this data set is beyond the level that our method can handle.

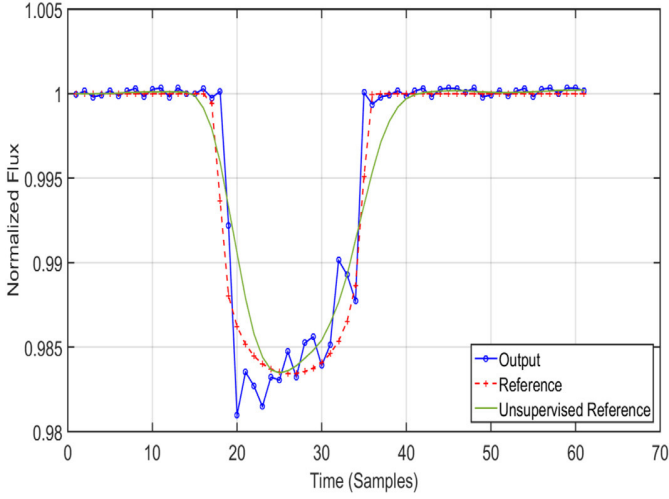
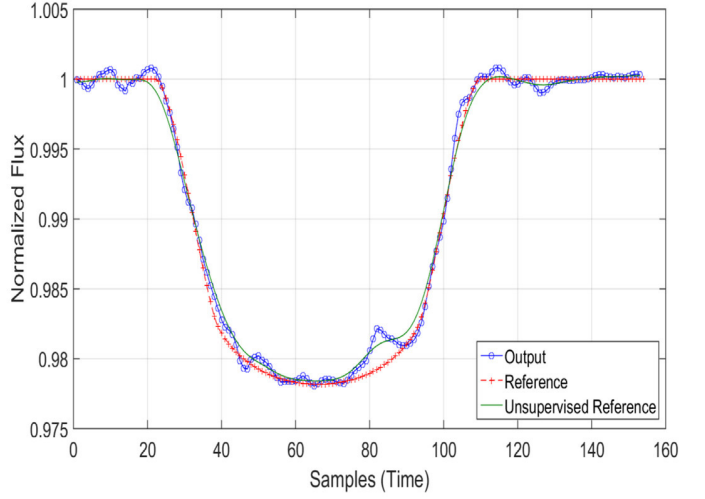

**Figure 8.** Output for FIR filter compared with unsupervised reference and fiducial reference for (a) GJ-1214 and (b) WASP-19b.
(A color version of this figure is available in the online journal.)

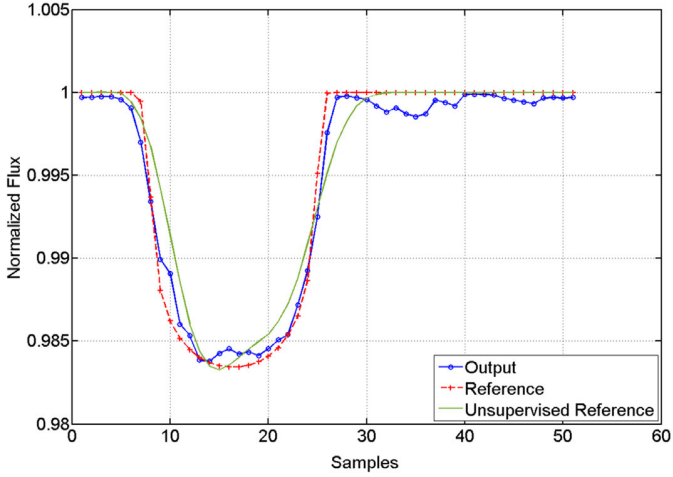
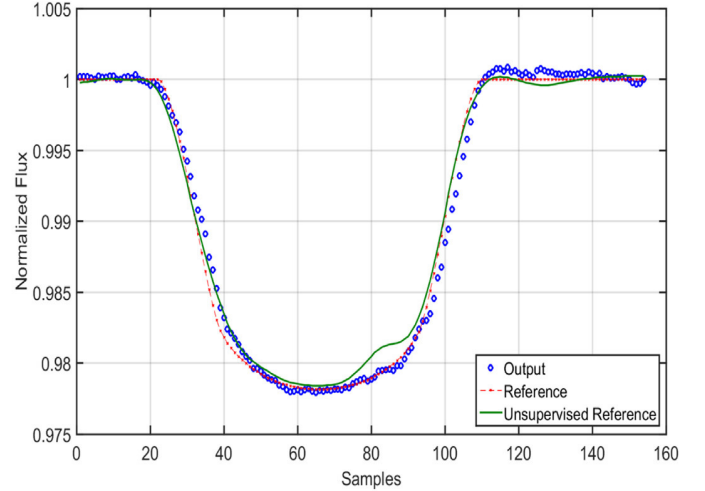

**Figure 9.** (a) NARX with clustering for GJ-1214 data set and (b) NARX with clustering for WASP-19b data set. The blue line represents the output average for every wavelength used in the simulation.
(A color version of this figure is available in the online journal.)

### 3.3. Synthetic Data

We constrain the effectiveness of our method versus different noise levels by using a a fiducial MA02 model with parameters taken from Table 2. We create 1000 data set for each of 43 different values of total S/N between 1 and $10^8$ (Figure 13). We considered both red and with noise to reach the required S/N with equal power. Each of these data sets where first run through JKTEBOP's task 9 and then computed an average value for the 1sigma uncertainty per each S/N value (black error bars in Figure 13). Additionally, these same data sets were first processed through our method, then run with JKTEBOP's task 9, and then computed and average value for the 1sigma uncertainty per S/N value (red error bars in Figure 13).

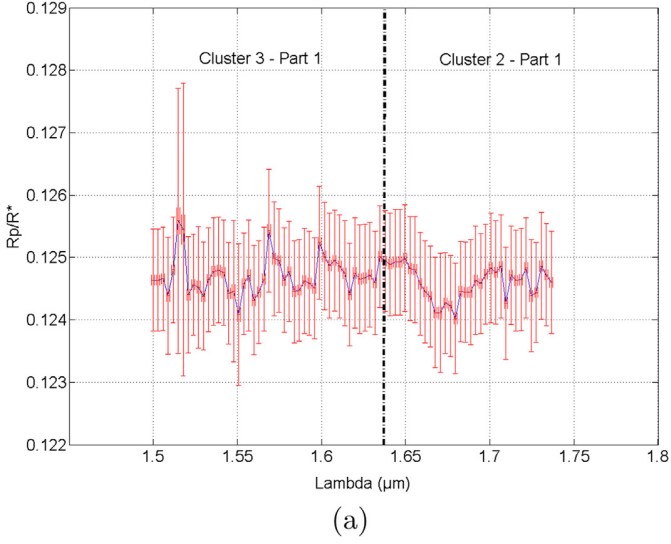 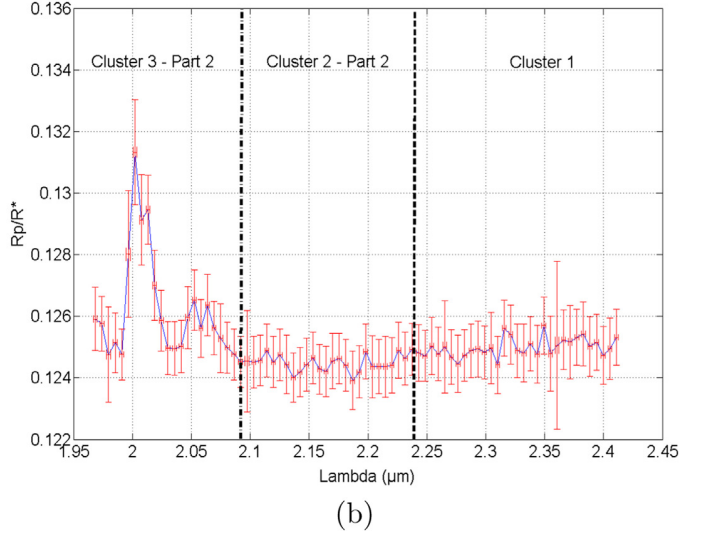

**Figure 10.** Radii ratio for (a) H-band window and (b) K-band window for GJ-1214 data set. Note that all features in this plot are likely product of the low S/N of the data and should not be trusted (see Section 3.3).
(A color version of this figure is available in the online journal.)

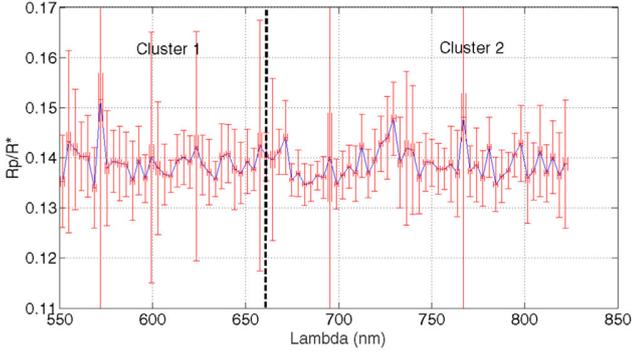

**Figure 11.** Radii ratio for WASP-19b data set.
(A color version of this figure is available in the online journal.)

As shown in Figure 13, the proposed improves the fit until $S/N \sim 10^2$. For $S/N < 10^2$, we eliminate the noise but the method adds an undesired offset. As shown in this figure, WASP-19b is in the correct range of noise for our method, but GJ-1214 is not in this range.

## 4. Conclusions

We proposed and verify a new method to pre-process transit light curves ahead of radius-ratio estimations to minimize systematic errors. The proposed method is NARX-based denoising system that was used to create an adaptive filter that reduces systematic errors that affect different wavelength channels similarly by the creation of an unsupervised reference created from the white-light data.

The core of our method is a NARX- neural network that requires a reference light curve, which we construct in an unsupervised fashion from the white light curve of each data set. Since this same reference is used for all channels, we are able to detect variation in radius-ratio within channels as required for exo-atmospheric characterization studies.

The method shows a visual improvement through each step, particularly, after using K-means clustering to create the reference signal. As shown in Figure 13, we validate our method and constrain its applicability regime testing synthetic data with different level of noise (white and red noise). We found that our method to be robust until $S/N > 10^2$.

We tested the method in two real data sets, one of good quality and one of low S/N, they behave as expected from our validation: providing a better and worst rest result than not using our method, respectively.

The authors acknowledge the financial support for the Project POSTDOC_DICYT, No 041613DA_POSTDOC, University of Santiago Center for multidisciplinary research on signal processing (Project Conicyt/ACT1120), Project USACH/Dicyt No. 061413SG, Program U-INICIA VID 2014, No. UI-02/2014, and Programa Enlace Fondecyt, University of Chile.

## Appendix A
## NARX Modeling

The NARX structure consists of many layers that contain one or more neurons. The first layer is called input layer. From the second layer, they are called hidden layer ("hidden" stands for differentiating from input and output layers, and has no other significance). We have a single hidden layer, but deep neural networks exists with multiple of these. Each input neuron is connected to every next neuron, called hidden neurons.

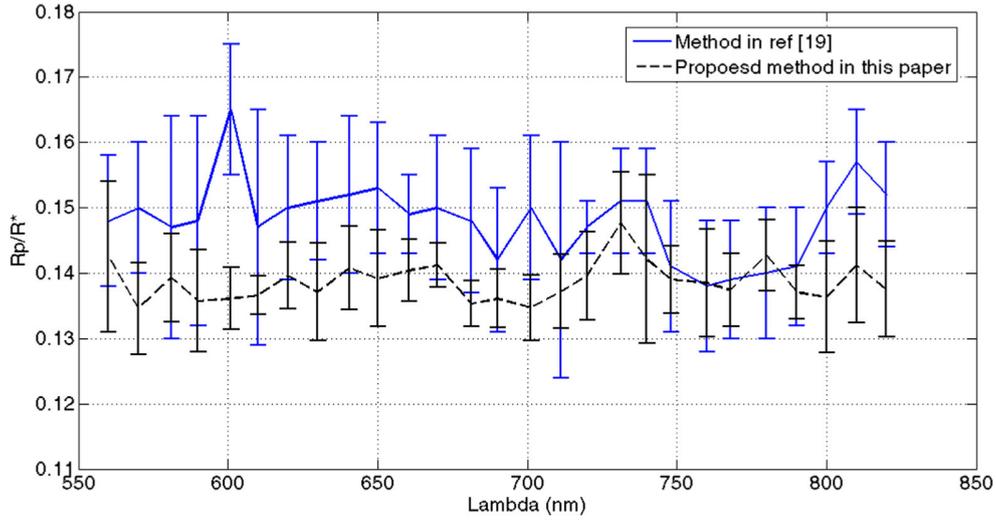

**Figure 12.** Radii ratio for proposed method for WASP-19b data set in comparison with (Sedaghati et al. 2015) that error bars were estimated by JKTEBOP software. (A color version of this figure is available in the online journal.)

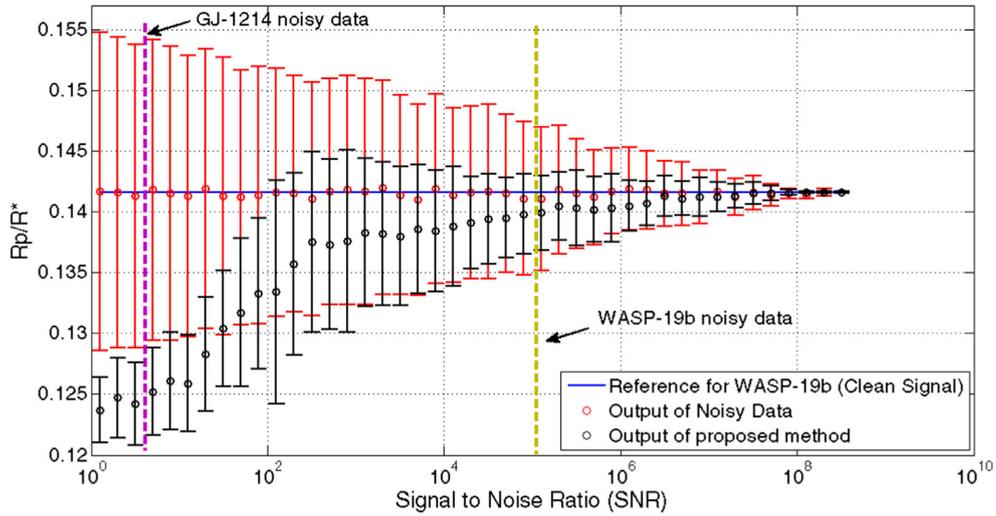

**Figure 13.** Radii ratio for proposed method for WASP-19b data set. (A color version of this figure is available in the online journal.)

In this work, a structure with three units (neurons) and six units is used in the hidden layer for WASP-19b, and GJ-1214 data set, respectively. This is determined by empirical testing on the data set, training several networks and estimating the generalization error of each. If there are too few hidden units (neurons for hidden layer), high training error and high generalization error due to under-fitting and high statistical bias will appear. If there are many hidden units, there will be low training error but still have high generalization error due to over-fitting and high variance.

In particular, it's not possible to sum up the design process for the neurons with a few simple rules of thumb. Instead, neural networks researchers have developed many design heuristics for the neurons, which help people get the behavior they want out of their nets. For example, such heuristics can be used to help determine how to trade off the number of neurons against the time required to train the network. The structure of NARX for WASP-19b, and GJ-1214 data set are shown in Figures 14 and 15 respectively.

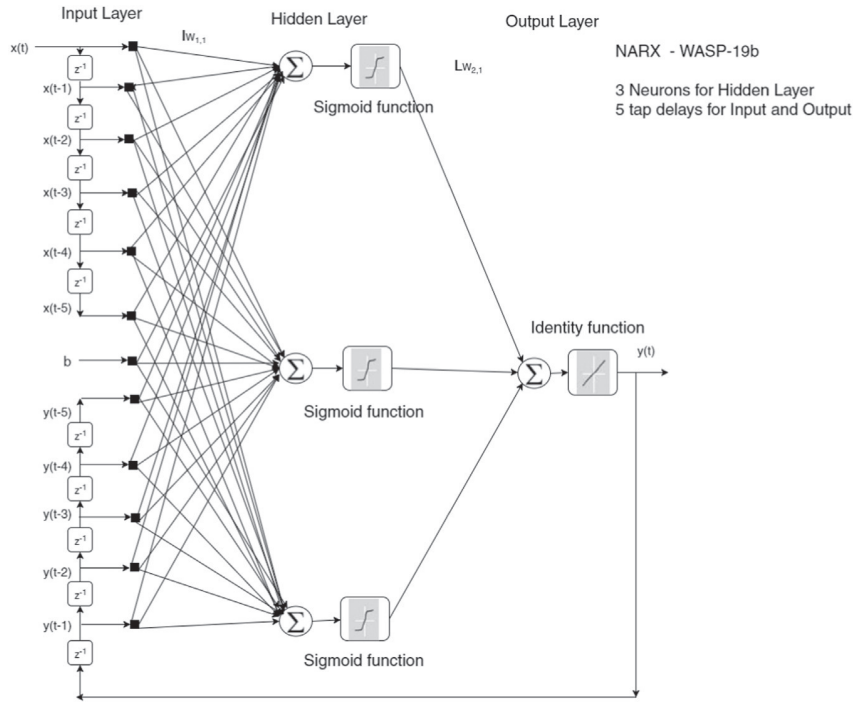

**Figure 14.** Structure for NARX for WASP-19b.

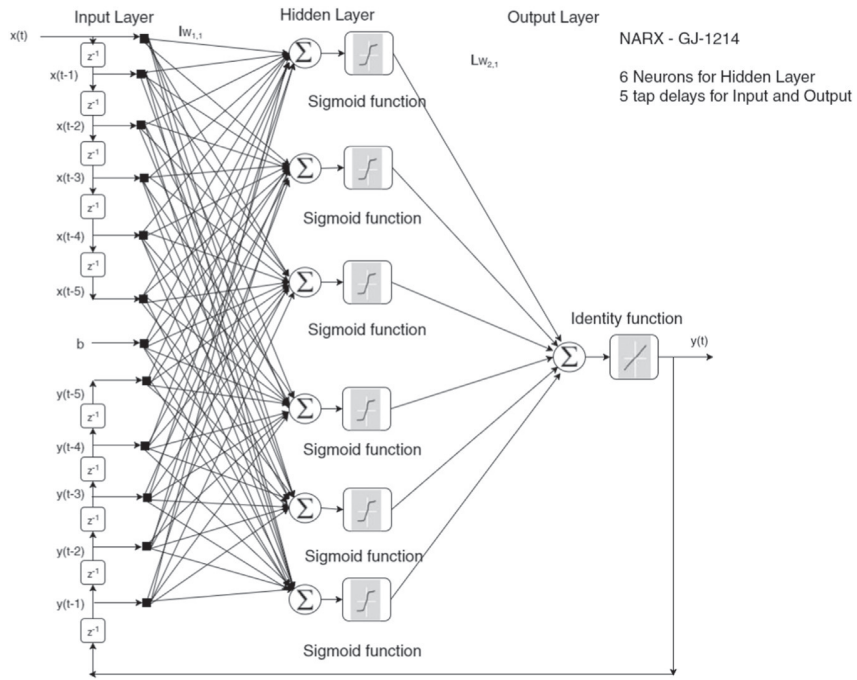

**Figure 15.** Structure for NARX for GJ-1214.

The NARX model is defined as

$$y(t) = [y(t-1), y(t-2),...,y(t-n_y),$$
$$(t), x(t-1), x(t-2),$$
$$..., x(T-n_x)] + \varepsilon(t), \quad (6)$$

where $x$ is a vector of time samples for each wavelength and $y$ is the denoised signal at the output of the NARX. The next value of the dependent output signal $y(t)$ is regressed on previous values of the output signal and previous values of an independent (exogenous) input signal. In this case, the reference curve found in step 3 for the exoplanet studied. The objective is to minimize the error $\varepsilon(t)$ in order to obtain a better approximation to the real value. The NARX thus computes a nonlinear dynamical function of its input. (Plett 2003)

The mathematical formulation of mapping performed by the network used in evaluation of the fitness function obtained by using appropriate matrix expressions and transfer functions for WASP-19b:

$$Iw_{1,1} = \begin{pmatrix} w_{1,1}^{(1)} & w_{1,2}^{(1)} & w_{1,3}^{(1)} \\ \vdots & \vdots & \vdots \\ w_{11,1}^{(1)} & w_{11,2}^{(1)} & w_{11,3}^{(1)} \end{pmatrix}, \quad (7)$$

$$Lw_{2,1} = \begin{pmatrix} w_{1,1}^{(2)} \\ w_{1,2}^{(2)} \\ w_{1,3}^{(2)} \end{pmatrix}, \quad (8)$$

and for GJ-1214:

$$Iw_{1,1} = \begin{pmatrix} w_{1,1}^{(1)} & ... & w_{1,6}^{(1)} \\ \vdots & \ddots & \vdots \\ w_{11,1}^{(1)} & ... & w_{11,6}^{(1)} \end{pmatrix}, \quad (9)$$

$$Lw_{2,1} = \begin{pmatrix} w_{1,1}^{(2)} \\ w_{1,2}^{(2)} \\ \vdots \\ w_{1,6}^{(2)} \end{pmatrix}, \quad (10)$$

where $Iw_{1,1}$ and $Lw_{2,1}$ are input layer and hidden layer weight matrix respectively. Also the output of NARX can be defined as:

$$y(t) = P_2(Lw_{2,1} . P_1(Iw_{1,1} . x(t) + b)), \quad (11)$$

where $P_1(t) = \frac{1}{1+e^{-t}}$ (Sigmoid function) and $P_2(t) = t$ (Identity function).

## Appendix B
## Unclustered Alternative

Figures 16 emphasizes the importance of clustering in the obtention of the radius-ratio, for both data sets while Figures 17 and 18 shows the result for each cluster, separately.

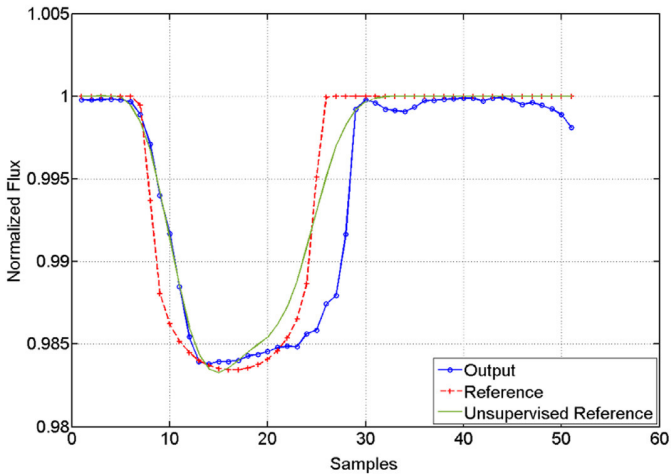 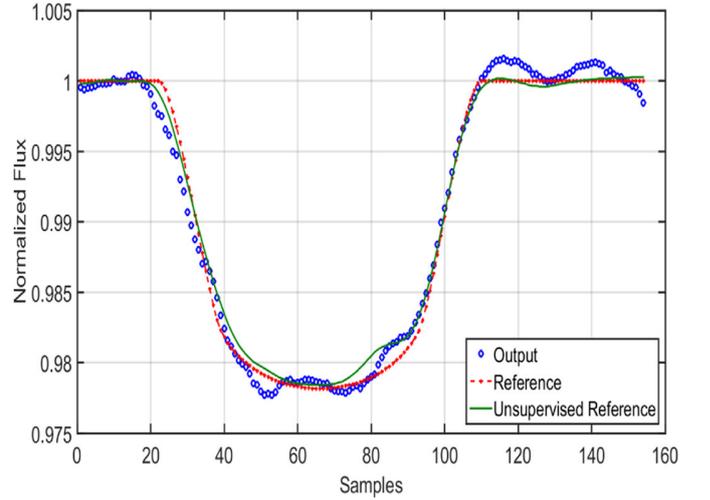

(a)        (b)

**Figure 16.** (a) Average output of the NARX over all wavelengths when building the unsupervised reference from all the wavelengths for GJ-1214 and (b) average output of the NARX over all wavelengths when building the unsupervised reference from the appropriate clusters for WASP-19b data set.
(A color version of this figure is available in the online journal.)

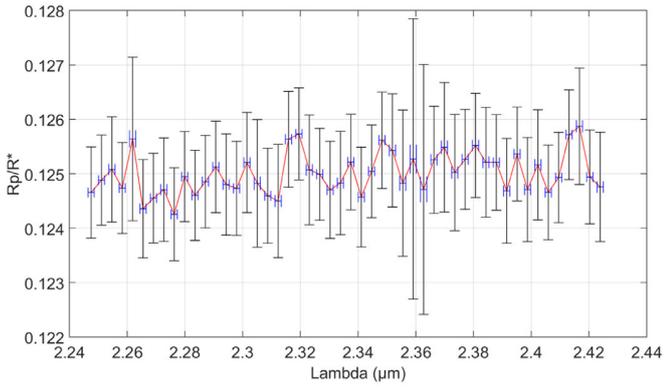
(a)

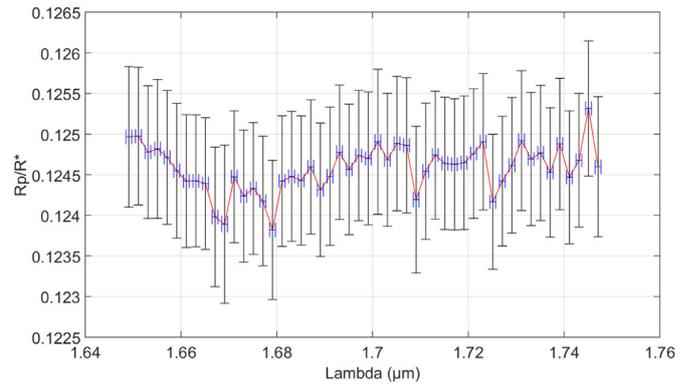
(b)

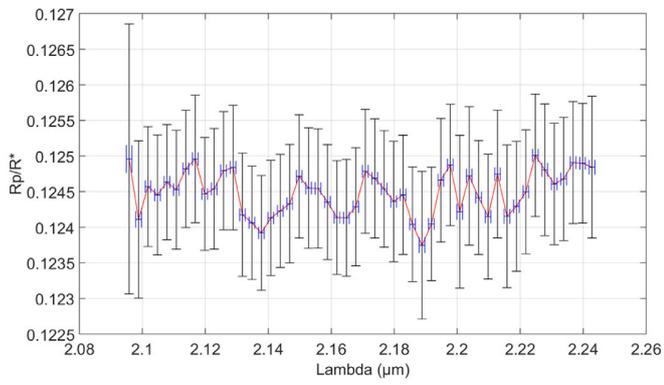
(c)

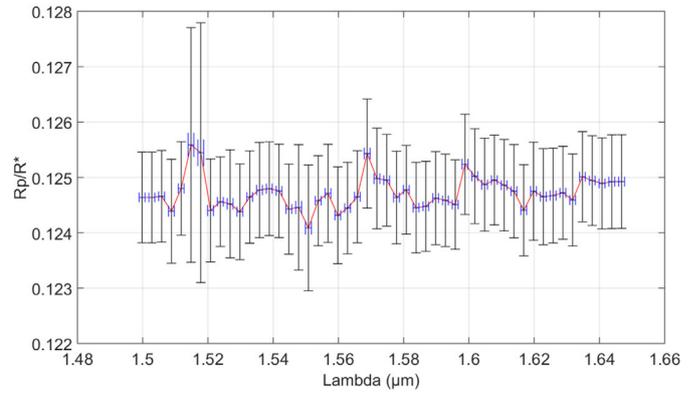
(d)

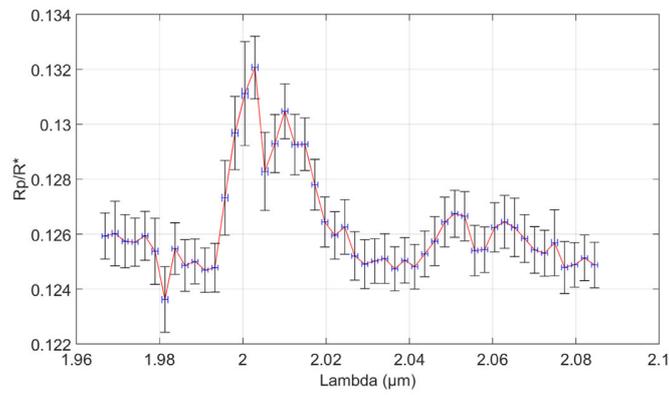
(e)

**Figure 17.** Radii ratio for Cluster 1, Cluster 21, Cluster 22, Cluster 31, Cluster 32, from top to bottom, respectively. GJ-1214 data set. (A color version of this figure is available in the online journal.)

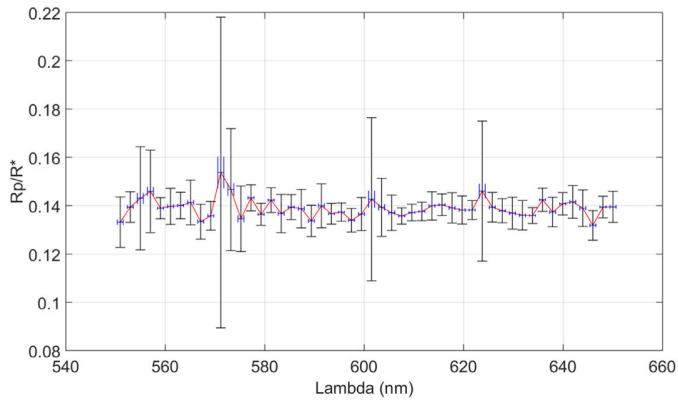 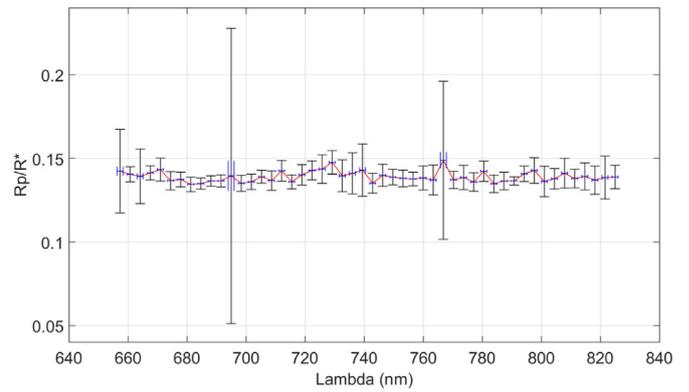

**Figure 18.** Radii ratio for (a) first and (b) second cluster for WASP-19b data set.
(A color version of this figure is available in the online journal.)